\documentclass[aps,prl,amsmath,amssymb,preprint]{revtex4}

\usepackage{float}
\usepackage{graphicx}
\usepackage{amsmath}
\usepackage{amssymb}
\usepackage{color}
\usepackage{soul}
\usepackage{subfigure}
\usepackage[sort&compress]{natbib}
\begin{document}
\title{Non-linear self-driven spectral tuning of Extreme Ultraviolet Femtosecond Pulses in monoatomic materials}

\author{Carino Ferrante$^{1,2,3*}$,
Emiliano Principi$^{4}$,
Andrea Marini$^{5}$,
Giovanni Batignani$^{3}$,
Giuseppe Fumero$^{3}$,
Alessandra Virga$^{2}$,
Laura Foglia$^4$,
Riccardo Mincigrucci$^4$,
Alberto Simoncig$^4$,
Carlo Spezzani$^4$,
Claudio Masciovecchio$^4$,
Tullio Scopigno$^{3,1*}$}

\affiliation{$^{1}$ Graphene Labs, Istituto Italiano di Tecnologia, Via Morego 30, 16163 Genova, Italy}
\affiliation{$^{2}$ Center for Life Nano Science @Sapienza, Istituto Italiano di Tecnologia, Viale Regina Elena 291, I-00161, Roma,Italy}
\affiliation{$^{3}$ Dipartimento di
	Fisica,~Universit\`a~di~Roma~\textquotedblleft La Sapienza", Piazzale Aldo Moro 5,~00185,~Roma,~Italy}

\affiliation{$^4$ Elettra-Sincrotrone Trieste S.p.C.A., SS 14-km 163.5, 34149 Basovizza, Trieste, Italy}
\affiliation{$^{5}$ Dipartimento di Scienze Fisiche e Chimiche, Universit\`a degli Studi dell’Aquila, Via Vetoio, 67100 L’Aquila, Italy}
\affiliation{$^{*}$ Corresponding authors}

\begin{abstract}
Self-action nonlinearity
is a key aspect -either as a foundational element or a detrimental factor- of several optical spectroscopies and photonic devices. Supercontinuum generation, wavelength converters and chirped pulse amplification are just a few examples.    
The recent advent of Free Electron Lasers (FEL) fostered building on nonlinearity to propose new concepts
and extend optical wavelengths paradigms
 for extreme ultraviolet (EUV) and X-ray regimes. No evidence for intrapulse dynamics, however, has been reported at such short wavelengths, where the light-matter interactions are ruled by the sharp absorption edges of core-electrons.
Here, we provide experimental evidence for self-phase modulation
 of femtosecond FEL pulses, which we exploit for fine self-driven spectral tunability by interaction with
sub-micrometric foils of selected monoatomic materials.  
Moving the pulse wavelength across the absorption edge, the spectral profile changes from a non-linear spectral blue-shift to a red-shifted broadening.
These findings are rationalized accounting for ultrafast ionization and delayed thermal response of highly excited 
electrons above and below threshold, respectively.
\end{abstract}
\maketitle

Progresses in non-linear optics have been facilitated by the development of 
high peak power, table-top pulsed laser sources operating in a relatively narrow 
spectral range spanning from the infrared to the ultraviolet.  
Such spectral constraint for fs-lasers restricts the interaction processes with condensed matter to those involving barely bound valence band electrons, limiting, in turn, the development of non-linear spectroscopies and photonic devices.
\\
A variety of non-linear interaction mechanisms occurs between light and matter in presence of intense radiation fields\cite{agrawal_2013}, which may alter the transient optical properties experienced by the light propagating in the material. The most elemental non-linear effect (NLE), arising at the single pulse level, is the self-induced spectral modulation, which produces the modification of the pulse spectral properties due to the polarizability induced in the matter by the pulse itself. Spectral modulation 
can originate from different optical phenomena, including Raman induced self-frequency shift, optical shock formation and self phase modulation (SPM) \cite{agrawal_2013}. 
 This latter, in particular, represents one of the primary tools used for tuning the spectral bandwidth at visible wavelengths by Kerr effect in transparent media. 
 Briefly, the presence of a strong laser field induces an intensity-dependent modulation of the material's refractive index which, in turn, produces a non-linear phase shift on the electromagnetic field, resulting in a broadening of its spectral profile\cite{agrawal_2013}.
 Used in combination with chirped
 mirrors, for example, SPM allows producing high intensity ultrashort optical pulses. On the other hand, it is detrimental in chirped pulse amplification\cite{strickland1985compression}, where it significantly distorts the recompressed pulse, limiting peak power and pulse contrast. Kerr effect also sets the temporal resolution limit in time resolved spectroscopies, due to pump-probe cross phase modulation\cite{doi:10.1063/1.372185,bat_acs}.

Exploring novel non-linear light-matter interactions in EUV and X-ray ranges, enabled by FEL and HHG sources, is recently attracting a lot of interest, based on the opportunity to design  new spectroscopic approaches\cite{Kowalewski2017,mukamel_optical,doi:10.1098/rsta.2017.0471} and to exploit NLEs in a high-energy regime\cite{nagler_turning_2009,glover2012x,PhysRevLett.111.233902,Shwartz2014,mincigrucci_role_2015,foglia2018,lam2018,Bencivengaeaaw5805,fidler_nonlinear_2019}. 
For example, theoretical and experimental efforts have been devoted to probe the effects of the core dynamics on the absorption properties of the sample, monitoring the photoinduced modification to the absorption spectra of materials pumped in the XUV and EUV ranges, either by the strong field of a single probe pulse \cite{mincigrucci_role_2015,nagler_turning_2009,PhysRevLett.123.163201} or by using an additional pump  pulse \cite{PhysRevLett.123.103001}. Differently, evidences for plasma induced SPM under \textit{valence} electrons photo-ionization in gas phase\cite{holzer2011femtosecond} or free carrier generation in transparent materials\cite{roy2013self,blanco2014controlling} have been reported under  visible excitation only: creation of plasmas by intense laser pulses and consequent laser-plasma interactions involve highly non-linear processes, producing a spectral blue-shift.
Critically, the lack of EUV and X-ray femtosecond pulses with controlled spectral and temporal profiles has prevented the observation of the self induced non-linear interactions with core electrons in non transparent materials.

Here, we demonstrate self-induced intrapulse dynamics of EUV 
seeded-FEL
 pulses 
propagating in sub-micrometer self-standing sample foils. 
Taking advantage of high peak power, 
transversal and longitudinal coherence and spectral stability of seeded-FEL sources\cite{PhysRevX.7.021043}, we 
exploit the NLEs in materials with metallic properties to
 alter the spectral shape of the transmitted FEL pulse. 
Interestingly, this phenomenon reveals a strong dependence on the interaction process between light and core electrons.
Specifically, we observed an asymmetric red-shifted spectral broadening when
exposing the sample to an intense EUV pulse with photon energy below a selected core absorption edge. 
This effect is rationalized by modelling SPM and delayed thermal response 
of electrons (DTRE)\cite{Marini_2013}, enabling the measurement of non-linear parameters of the sample material.
On the other hand, increasing the EUV photon energy a few eV above a targeted absorption edge, 
we measured a pronounced non-linear blue-shift, ascribable to a SPM effect induced by photo-induced core electron ionization. 

\begin{figure}
	\centering
	\includegraphics[width=15.5cm]{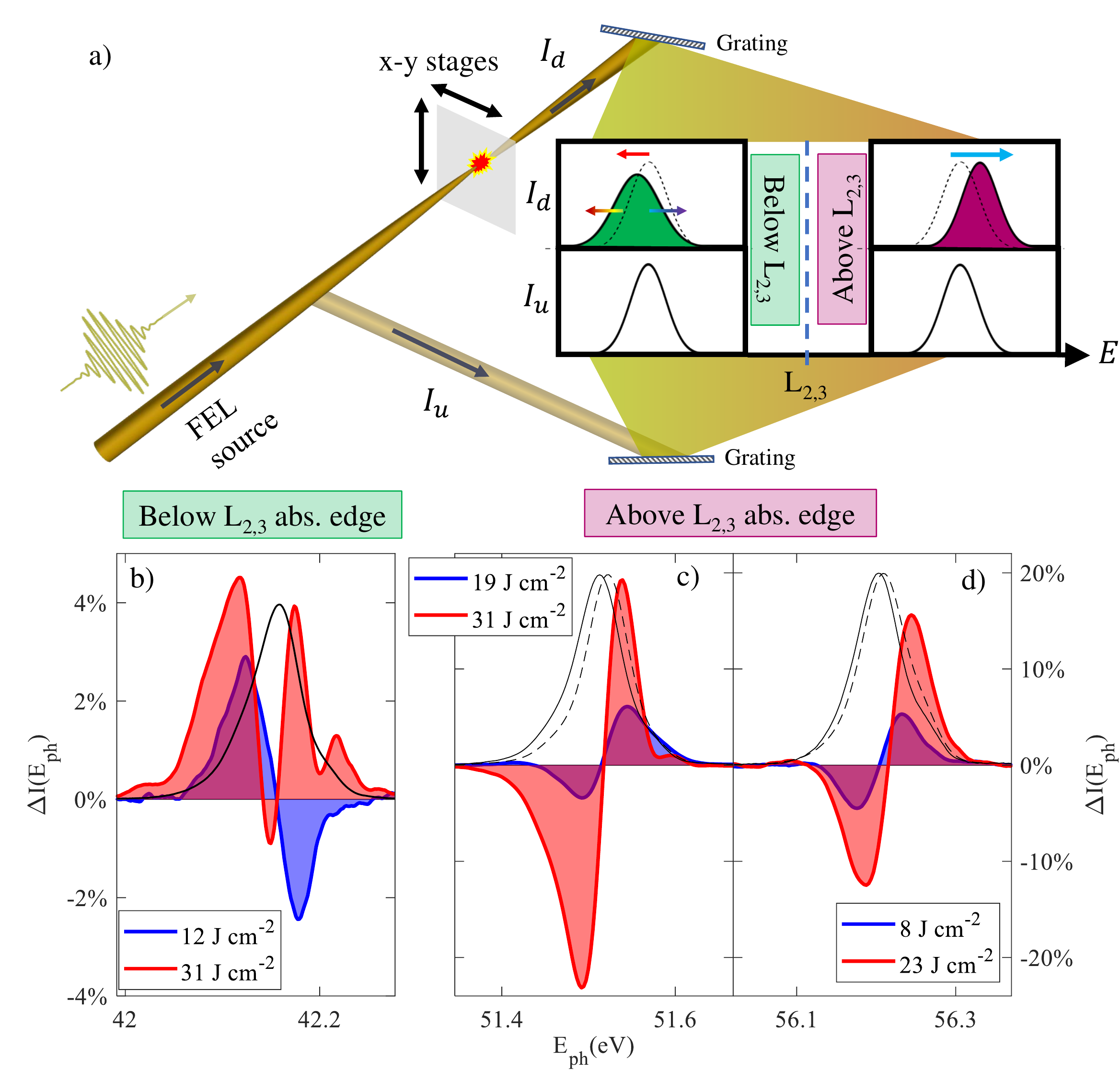} 
	\caption{\textbf{Experimental results in Mg.} a) Simplified sketch of the experimental setup (see Methods). The single-shot spectra of the pulses are collected with two spectrometers (depicted as a couple of gratings) upstream ($I_{\rm u}(E_{\rm ph})$) and downstream ($I_{\rm d}(E_{\rm ph})$) of a Mg sample foil.
	The NLEs are obtained comparing the two FEL spectra, demonstrating a spectral blue-shift above the $L_{2,3}$ absorption edge. On the contrary, spectral red-shift and broadening can be observed below edge. 
		 The averages of the differences $\Delta I=\langle I_{\rm d}-I_{\rm u}\rangle$ is reported for a FEL photon energy ($\sim 42.1$ eV)  below the Mg L$_{2,3}$-edge   (b) and for two photon energies ($\sim 51.5$ and $\sim 56.2$ eV) above it (c,d). The red and blue curves correspond to $\Delta I$ at the highest and lowest fluence.
		 Averaged FEL spectra $\langle I_{\rm u}(E_{\rm ph})\rangle$, scaled to improve readability, are shown by solid black lines.  The blue-shift in c,d can also be appreciated comparing $\langle I_{\rm u}(E_{\rm ph})\rangle$ and $\langle I_{\rm d}(E_{\rm ph})\rangle$ at the largest fluence (dashed black lines).
	\label{fig_experiment}
	} 	
\end{figure}

The experiment, sketched in Fig. \ref{fig_experiment}a, is carried out at the beamline EIS-TIMEX 
of the FERMI FEL in Trieste (Italy)\cite{Masciovecchio:xj5007}. FERMI delivers intense nearly
transform-limited sub-100 fs pulses, finely tunable in wavelength and intensity 
(see Methods for details).
The pulses are focused  on free-standing sub-micrometric (100-300 nm) 
sample foils of metals  and semiconductors, mounted on supporting rings.
Particularly, measurements at different photon energies in the range 32-56 eV and different fluences
are performed on Mg and Se that exhibit sharp absorption edges 
($L_{2,3}$ $\sim 50$ eV  and $M_{4,5}$ $\sim 55$ eV, respectively\cite{thompson2001x}) 
in the experimental spectral window.
Every single FEL pulse is monitored by two spectrometers. The first one, located upstream of the experimental chamber, measures the FEL spectral intensity ($I_{\rm u}(E_{\rm ph})$) before the interaction with the sample. The second one, placed downstream of the chamber, performs an analogous measurement for the light transmitted by the sample  ($I_{\rm d}(E_{\rm ph})$). In front of this latter spectrometer a rotating wheel with different metallic filters  preserves the light sensors from saturation.
The averages of differential single-shot normalized spectra 
($\Delta I(E_{\rm ph})=\langle I_{\rm d}(E_{\rm ph})-I_{\rm u}(E_{\rm ph})\rangle$, see Methods) for a 140 nm thick 
Mg self-standing foil, protected on both sides by an Al coating (19 nm)\cite{mincigrucci_role_2015}, is reported in 
Fig. \ref{fig_experiment}b,c,d. 
 
\begin{figure}
	\centering
	\includegraphics[width=10cm]{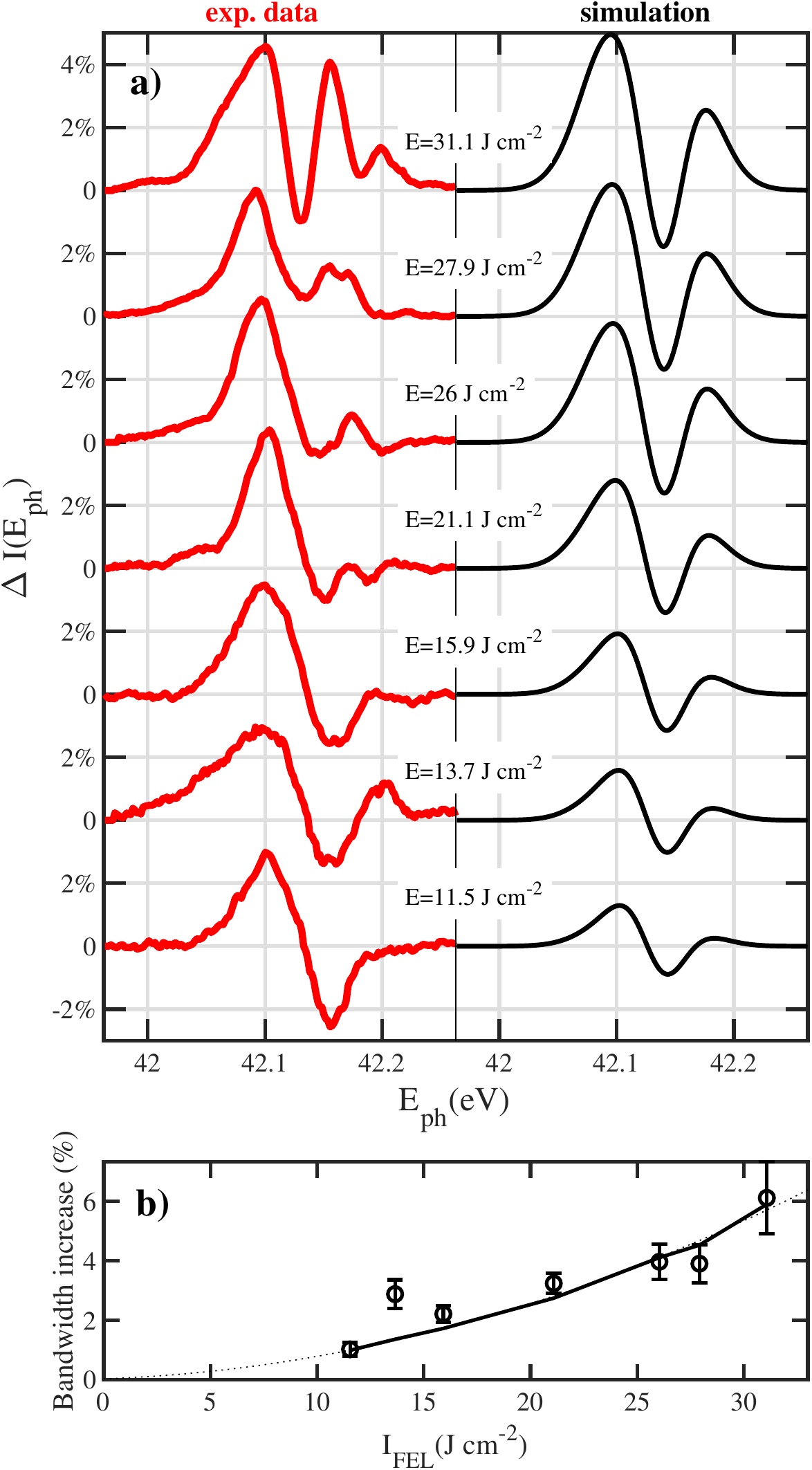} 
	\caption{\textbf{Theoretical simulation of experimental results below the absorption edge.} 
		a) The experimental data (red lines) reported for different FEL fluences in $140$ nm of Mg are compared with the relative simulation (black lines) of DTRE and SPM NLEs obtained from a fitting procedure
		of Eq. \ref{GNLSE}.
		 The modification of  pulse intensity ($\Delta I= |\psi(L,E_{\rm ph})|^2- |\psi(0,E_{\rm ph})|^2$) is calculated (see Methods).  b) The percentage increases of bandwidth, observed in our measurements (circles) and in the  theoretical curves (solid line), are depicted.
The dotted line is a parabolic guide for eye.
	} 	\label{fig_compare}
\end{figure}

Notably, the $\Delta I$ signal in Mg reveals a different system response 
upon light-excitation above and below $L_{2,3}$ absorption edge.
As shown by the raw spectra in Fig. \ref{fig_experiment}c,d ($\langle I_{\rm u}(E_{\rm ph})\rangle$, solid black line,  and $\langle I_{\rm d}(E_{\rm ph})\rangle$, dashed black line),  a
  blue-shift dominates the non-linear spectral modification above the edge. Accordingly, $\Delta I(E_{ph})$  (colored blue and red areas) shows a dispersive lineshape. This is opposed to the case of gas photo-ionization in the visible, where  there exists an initial propagation regime in which plasma generation is negligible and Kerr SPM dominates\cite{holzer2011femtosecond}. 
Below the $L_{2,3}$-edge, the spectral shape of $\Delta I(E_{\rm ph})$ is more complex and fluence dependent, 
as reported in Fig. \ref{fig_experiment}b.
Specifically, at the highest FEL fluence, $\Delta I(E_{\rm ph})$ exhibits two positive lobes, 
signature of a spectral broadening (quantified in Fig. \ref{fig_compare}b).
Upon decreasing the FEL fluence, $\Delta I(E_{\rm ph})$ gradually evolves to a dispersive shape, 
corresponding to a spectral red-shift.

We ascribe the observed dependence of the spectral behavior on the photon frequency to the presence of a core absorption edge in the explored spectral region, drastically altering the interaction of EUV photons with the material. 
In order to rationalize the measured spectral shifts, we model the data by describing optical pulse propagation in the non-linear medium. 
When an optical field 
${\bf E}({\bf r},t) = {\rm Re} [\psi({\bf r},t) {\rm e}^{ik_0z-i\omega_0t}\hat{\bf n}]$ 
with generic complex polarization unit vector $\hat{\bf n}$, optical envelope $\psi({\bf r},t)$, 
carrier angular frequency $\omega_0$ and wave-vector $k_0 = \omega_0 n(\omega_0)/c$ 
propagates in a non-linear medium with linear refractive index $n(\omega_0)$, it undergoes a self-induced phase 
shift\cite{agrawal_2013} quantified as
\begin{equation}
\phi_{\rm NL}({\bf r},t)=\chi^{(3)} |\psi({\bf r},t)|^2 k_0 L,
\label{eq:SPM_th}
\end{equation}
where $L$ is the sample thickness, and $\chi^{(3)}$ is the third-order Kerr susceptibility, which represents
the  light capability to modify the refractive index of the absorbing material. 
This time-dependent non-linear phase shift, induced by a non-linear polarization (${\bf P}_{\rm NL}({\bf r},t)$), 
leads to self-induced spectral broadening of ultrashort pulses \cite{alfano1970}.
As shown in Eq. \ref{eq:SPM_th}, Kerr-induced SPM is proportional to the $\chi^{(3)}$ of the material. 
Owing to resonant interband electron dynamics, in the visible range metals exhibit higher 
$\chi^{(3)}$ values ($\sim10^{-16}$ m$^2$V$^{-2}$) than semiconductors 
($2.8\times10^{-18}$ m$^2$V$^{-2}$ in silicon), glasses ($\sim10^{-22}$ m$^2$V$^{-2}$) and 
solvents ($\sim10^{-20}$ m$^2$V$^{-2}$) \cite{boyd2019nonlinear}. 
Such highly resonant non-linear behavior
 in metals cannot be efficiently exploited
in the visible range, as
 it is accompanied by poor light transmission with the noteworthy exception of FWM in graphene\cite{PhysRevLett.105.097401,malard_graphene, virga_coherent_2019}.
 This forces resorting to surface plasmon polariton waves to obtain non-linear functionalities in metal-based table-top photonic systems\cite{kauranen2012nonlinear}. 
EUV and X-ray photons are instead well transmitted by submicrometric metallic media and can
excite core photoelectrons, potentially representing the most suitable platform for full exploitation of NLEs in metals. 

In our experiment, when the EUV FEL photon energy slightly exceeds selected core electron 
binding energies (above-edge condition), core photoelectrons
are massively promoted nearly above the Fermi level, possibly leading to saturable absorption\cite{mincigrucci_role_2015}. 
In this condition, photoelectrons form a transient hot dense ionized 
plasma, which gives rise to a SPM different than that originated by Kerr effects in transparent media.
In the case of free electron plasma generation driven by pulse absorption, 
the charges induce a sudden decrease of the refractive index in the system  \cite{penetrante1992ionization,baudisch2018ultrafast}. 
Such an absorption-driven effect induces a non-linear modification of the light propagation, accelerating the trailing edge of the pulse and consequently blue-shifting the spectrum. 
Specifically, the leading edge of the pulse propagates through an absorbing medium, while the trailing part, due to the free-electron dispersion produced by the leading front, experiences a reduced refractive index and is hence accelerated. As a result of such asymmetric temporal compression the pulse spectral components are blue-shifted, which rationalizes the experimental results in fig. \ref{fig_experiment}c,d. 

Below the absorption edge, the sample absorption coefficient for EUV photons is radically lower 
and the electron plasma density is negligible, leading to the suppression of the intense spectral blue-shift. 
Nevertheless,  spectral NLEs, induced 
by the high EUV local electric field in the excited sample volume, can still be observed. 
Specifically, a clear spectral broadening  
 (red curves in Fig. \ref{fig_experiment}b) is obtained at high FEL fluence. Such an effect can be rationalized  in terms of Kerr-induced SPM under non absorbing conditions. However, SPM alone cannot account for the 
spectral red-shift occurring at low fluence (green curves in Fig. \ref{fig_compare_only}), as demonstrated by simulation in Fig. \ref{fig_compare_only}. Under low local electric field excitation, spectral red-shifts can be explained in terms of DTRE\cite{Marini_2013,marini_faraday}.
DTRE is frequently observed in plasmas, doped semiconductors and metals, because optical interaction with electrons leads to ultrafast out-of-equilibrium electronic heating, which modulates the refractive index with a delayed response, arising from the excitation of hot conduction electrons ($\tau_{\rm th}=1$ fs) and its consequent relaxation to equilibrium via electron-phonon scattering ($\tau_{\rm r}=100$ fs)\cite{cai_optical_2010,PhysRevB.87.035139}. 
Owing to DTRE, the temporal relaxation dynamics  is accompanied by a non-linear spectral red-shift in the frequency domain\cite{Marini_2013}. 

Such a complex ultrafast non-linear dynamics, 
triggered by the concomitant SPM and DTRE effects, can be phenomenologically modeled  through 
a two-temperature model \cite{Marini_2013,marini_faraday}, which determines the following time-dependent 
non-linear polarization (see Methods):
\begin{equation}
{\bf P}_{\rm NL}({\bf r},t)= \epsilon_0 {\rm Re} \left\{ \chi^{(3)} \left[(1-f_{\rm T})|\psi({\bf r},t)|^2 + f_{\rm T} \int_0^{\infty}dt'h_{\rm T}(t')|\psi({\bf r},t-t')|^2\right] \psi({\bf r},t){\rm e}^{ik_0z-i\omega_0t}\hat{\bf n} \right\}, \label{PolEq}
\end{equation}
where $f_{\rm T}$ is the thermal fraction and $h_{\rm T}(t) = (\tau_{\rm th}-\tau_{\rm r})^{-1}\left(e^{-t/\tau_{\rm th}}-e^{-t/\tau_{\rm r}}\right)$ is the thermal response function.  
In Fig. \ref{fig_compare}a we report the spectral modifications induced by sample nonlinearity, for selected fluence values. Experimental data are best represented by simulations obtained propagating a 59 fs transform limited pulse in a material with $f_{\rm T}=0.29$, and $\chi^{(3)}=1.45\cdot 10^{-23}$ m$^2$V$^{-2}$ and carrier photon energy of 42.12 eV.
 
Notably, the $\chi^{(3)}$ value is larger than measured in Si$_3$N$_4$ by transient 
FWM\cite{foglia2018} ($6\cdot10^{-24}$ m$^2$V$^{-2}$), indicating stronger NLE in metals.
Moreover, the DTRE effect, related to the interaction with free electrons, has a tangible role below the absorption edges, as expected for metals in the visible range.
The extra oscillations observed in the experimental data at $\sim$ 42.2 eV can be ascribed to spectral deviation of the FEL source with respect to an ideal transform limited Gaussian pulse, as discussed in  supplementary information.
As shown in Fig. \ref{fig_compare}, the role of DTRE contribution is more evident in the energy region up to 16 J cm$^{-2}$. The measured $f_{\rm T}$ implies that thermal excitation of electrons is comparable in efficiency to the instantaneous Kerr nonlinearity. 
In addition, our results indicate that hot electron thermalization occurs in the fs timescale
and consequently collision-induced dephasing plays a major role in the electron dynamics, 
thus preventing coherent effects like Rabi oscillations or self-induced transparency\cite{PhysRevLett.110.243901}. 

\begin{figure}
	\centering
	\includegraphics[width=10cm]{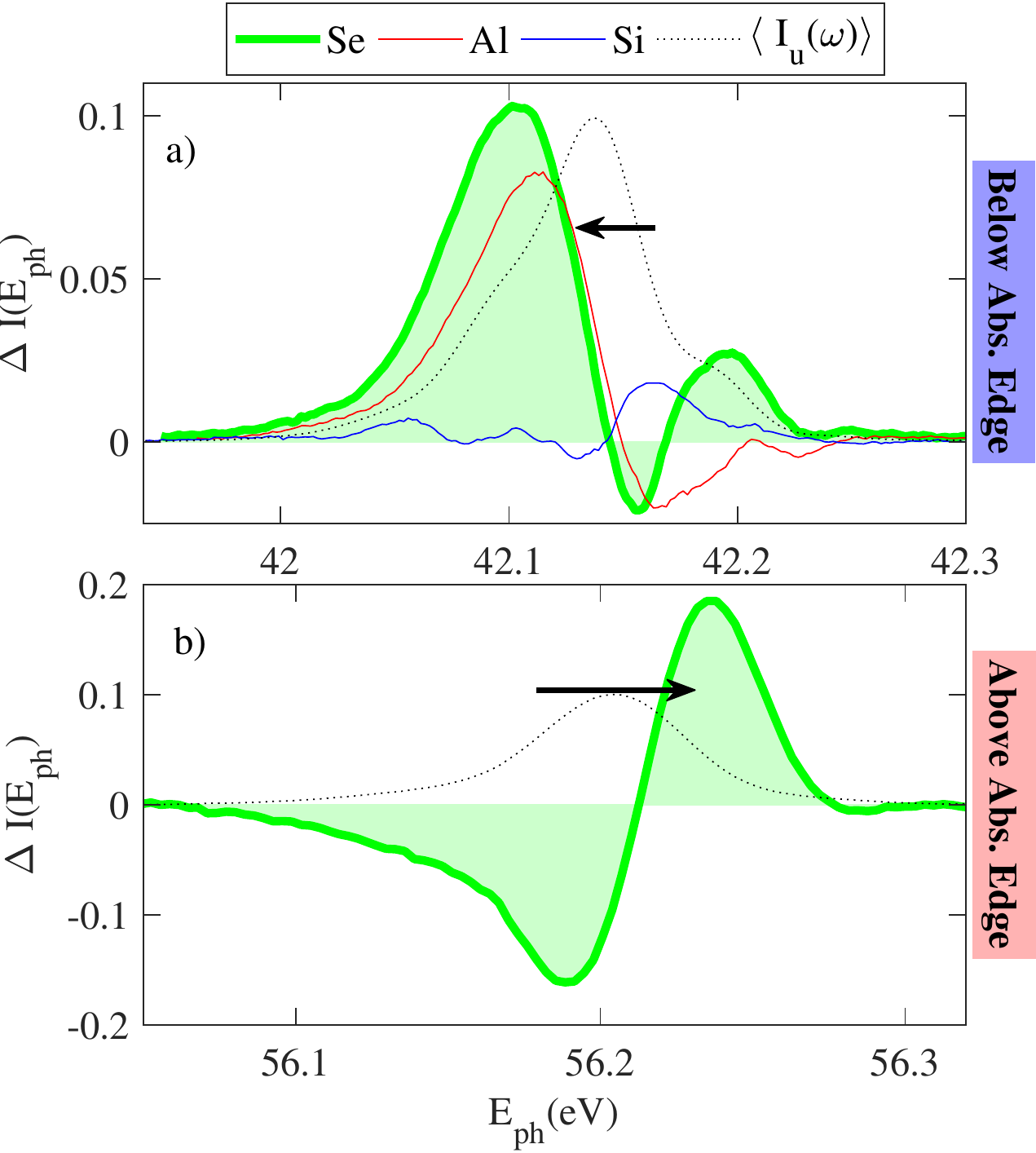} 
	\caption{\textbf{Spectral modification in sub-micrometric films.} a) The difference of normalized spectra before and after the sample interaction in a 300 nm  Al film  (red line), 100 nm of Si (blue line) and 120 nm of Se (green area) at 42.1e eV with a fluence of 31 J cm$^{-2}$. b) The measurement  is performed in Se film above the M$_{4,5}$ absorption edge (at 55 eV) with a fluence of 9 J cm$^{-2}$. 
		$\langle I_{\rm u}(E_{\rm ph})\rangle$ of the sample, divided by a factor 10, is depicted by the black dotted lines.
		The horizontal black arrows indicate the shift direction.
	} 	\label{fig_materials}
\end{figure} 

Similar measurements are carried out at $\sim42.1$ eV for other sub-micrometric sample foils, in order to extend the experimental investigation to a wider class of materials. 
In particular, in Fig. \ref{fig_materials}a we report $\Delta I$ measured in Al (300 nm), 
Si (100 nm) and Se (120 nm).
In agreement with the Mg case, since the experimental photon energy is below the core binding energy of the three materials, all the measurements do not show any spectral blue-shift. 
Al and Se exhibit a spectral modification, while no substantial effect is observed in Si, confirming the higher $\chi^{(3)}$ of metallic samples also in EUV regime, as previously suggested by the comparison with transient FWM experiment\cite{foglia2018}.
Furthermore, measurements in Se above the M$_{4,5}$ absorption edge (see Fig. \ref{fig_materials}b) 
provide an additional evidence of plasma induced spectral blue-shift. 
Further improvement of the NLEs efficiency can be achieved using thicker samples and photo-exciting below the absorption edge. In particular, the extracted Mg $\chi^{(3)}$ indicates that a micrometer sample
can generate an increase of the FEL full width half maximum up to a 2.5 factor (0.14 eV).

In summary, we have explored the sub-picosecond non-linear response of Mg and Se in the extreme ultraviolet and, focusing high fluence EUV beam into opaque sub-micrometric foils, we have shown how to exploit NLEs for driving a self-induced modification of the pulse spectrum.
Specifically, we reveal a plasma induced blue-shift when the FEL pulse photon energy exceeds core electrons binding energy. In striking contrast, when core electron photo-absorption is not activated, two different regimes are observed: a red-shift at low fluences ($<16$ J cm$^{-2}$), and a spectral broadening at higher fluences.
This intensity dependence is rationalized accounting for the concurring presence of two different non-linear processes, namely Kerr-induced SPM and DTRE, allowing to measure the thermal fraction, $\sim$30\%  in Mg, and the third-order nonlinearity, $\chi_{\rm Mg}^{(3)}=1.45\cdot 10^{-23}$ m$^2$V$^{-2}$.
The presented results, verified in different materials, provide the first evidence of self-induced spectral beam modification in the EUV, indicating the crucial role of the core electron absorption edges for such NLEs and demonstrating the higher non-linear efficiency in metallic samples, compared to semiconducting materials.
We anticipate that the full exploitation of these NLEs will lead to a fine control of the EUV ultrashort light pulse spectral profile in FEL facilities or in table-top HHG setups, creating new opportunities in non-linear photonics and time-resolved spectroscopies\cite{Kowalewski2017,doi:10.1098/rsta.2017.0471,bressler2004ultrafast}.

\section{Methods}
\subsection{Experimental setup}
 The FERMI FEL source is optimized to deliver nearly-transform limited EUV pulses with an intensity of 
  $\sim75$ $\mu$J, measured by a calibrated N\textsubscript{2} ionization chamber. 
 The system is able to emit photons from 12 eV to 62 eV, with the pulse duration that depends on the photon energy. In our experiment, they are 42.1, 51.5  and  56.2 eV  and 59, 55 and 54 fs, respectively\cite{PhysRevX.7.021043}.
 The FEL beam is delivered to the 
 end-station using grazing incidence (about 2 degrees) carbon coated mirrors.
 The experimental setup consists of two EUV single-shot spectrometers  positioned upstream (PRESTO) and
 downstream (WEST) of an experimental chamber where the sample is accommodated. The resolving power of the two spectrometers is about $10^4$, appropriate to monitor the FEL peak profiles with typical bandwidth of about 0.03 nm. 
  No monochromators are involved in the setup.   
 The FEL beam is focused on the sample by a gold coated ellipsoidal mirror with a focal length of 1400 mm.
 This tight focusing in a region of diameter $\sim11 \mu$m permits reaching fluences of about $\sim31$ J cm$^{-2}$.
 Absorption of intense FEL pulses induces a permanent damage\cite{mincigrucci_role_2015} on samples, 
 therefore measurements have been carried out in single shot fashion. 
 In the probed spectral region, the above threshold absorption is practically constant (less than 2\%). Therefore, it does not contribute to spectral distortions, which can only be ascribed to non-linear effects.
 After each FEL shot, selected by a mechanical fast shutter, the sample is moved to an unexposed region, using an  automatically moving  5-axes manipulator
 preserving both the focus and the normal incidence of the FEL beam.
  Single FEL shots are selected with a repetition rate of about 0.25 Hz by a mechanical fast shutter with a diamater of 6 mm positioned between the sample and the ellipsoidal mirror.
  The normalized spectra before  $I_{\rm u}(E_{\rm ph})$ and after $I_{\rm d}(E_{\rm ph})$ the sample are collected for each single FEL shot. Preliminary, $I_{\rm u}(E_{\rm ph})$ and $I_{\rm d}(E_{\rm ph})$ are both energy shifted to maximise the spectral overlap of different shots at the upstream detector, which is then used as a reference: in fact, self-induced spectral modifications are obtained averaging single-shot spectral differences $\langle I_{\rm d}(E_{\rm ph})-I_{\rm u}(E_{\rm ph})\rangle$ over $\sim 100$ FEL shots.
   
 To mitigate artifacts from the long-term fluctuations of the spectrometers’ calibration, 
we collected short sequences (1-2 minutes) of calibration and measurement data  subsequently.
 The averaged difference between the simultaneous normalized measurements of the two spectrometers, reported in Figs. \ref{fig_experiment},\ref{fig_materials}, have been corrected by subtracting the calibration spectrum.
\subsection{Theoretical calculations}
For photon energies below the ionization threshold, in order to account for the ultrafast dynamics of Mg, we adopt a phenomenological two-temperature model \cite{Marini_2013,marini_faraday} 
\begin{eqnarray}
&& \dot{{\cal N}} = - \tau_{\rm th}^{-1}{\cal N} + P_{\rm A}, \\
&& \dot{T}_{\rm e} = \tau_{\rm r}^{-1}(T_{\rm eq} - T_{\rm e}) + (\gamma_{\rm e}/C_{\rm e}) {\cal N},
\end{eqnarray}
describing the temporal evolution of non-thermalized electrons with energy density ${\cal N}({\bf r},t)$ thermalizing at rate $\tau_{\rm th}^{-1}$ to the out-of-equilibrium temperature $T_{\rm e}({\bf r},t)$ and subsequent relaxation to the equilibrium lattice temperature $T_{\rm eq}$ at rate $\tau_{\rm r}^{-1}$ upon radiative excitation by an electromagnetic pulse with electric field ${\bf E}({\bf r},t) = {\rm Re} [\psi({\bf r},t) {\rm e}^{ik_0z-i\omega_0t}\hat{\bf n}]$ , where $\psi({\bf r},t)$ is the optical envelope, $\omega_0$ is the carrier angular frequency, $k_0 = \omega_0/c$ is the carrier wavevector, and $\hat{\bf n}$ is the polarization unit vector. The absorbed power per unit volume averaged over fast temporal oscillations is $P_{\rm A} ({\bf r},t) = (1/2)\epsilon_0\epsilon''(\omega_0)\omega_0|\psi|^2$, where $\epsilon(\omega_0)$ is the metal linear dielectric constant at the carrier angular frequency and the double prime indicates the imaginary part, while $\gamma_{\rm e}$ represents the electron-electron collision rate and $C_{\rm e}$ is the electron heat capacity per unit volume. Considering the previous equation, the absorbed power $P_{\rm A} ({\bf r},t)$ depends on time over the timescale of the radiation envelope (59 fs).
The two-temperature model is derived from first-principles by the method of moments starting directly from the Boltzmann equations for lattice and free electron fluids in the relaxation approximation (non-ideal plasma, electron-electron and electron-phonon collisions are phenomenologically accounted for by the relaxation rates)\cite{Marini_2013}.
Such model can be analytically solved for arbitrary pulses by Fourier transform, leading to the temporal evolution of the electron temperature variation 
\begin{equation}
\Delta T_{\rm e}(t) = T_{\rm e}(t) - T_{\rm eq} = C_{\rm e}^{-1}\gamma_{\rm e}\tau_{\rm r}\tau_{\rm th}\int_0^{\infty}dt'h_{\rm T}(t')P_{\rm A}(t-t'),
\end{equation}
where $h_{\rm T}(t) = (\tau_{\rm r}-\tau_{\rm th})^{-1}\left(e^{-t/\tau_{\rm r}}-e^{-t/\tau_{\rm th}}\right)$ is the thermal response function, 
$\tau_{\rm th}$ (of the order of fs) is the electron thermalization time arising from collisions that at the same time produce electronic dephasing. After ultrafast heating, the increased electron temperature $T_{\rm e}$ relaxes to the lattice temperature with the characteristic time $\tau_{\rm r}$, accounting for electron-phonon collisions.
The main limitation of the model lies in its inadequacy to produce meaningful predictions for radiation pulses with time duration shorter than $\tau_{\rm th}$ because in such regime the electron distribution is utterly non-thermal and the definition of an out-of-equilibrium hot electron temperature is meaningless. However, in our present investigation, the radiation pulse duration is longer than $\tau_{\rm th}$ and hence the model is fully adequate to produce accurate predictions. Further, the model is spatially local, assumes a homogeneous lattice, and neglects hot-electrons diffusion occurring over timescales longer than our pulse duration \cite{block2019tracking,najafi2017super}.

The ultrafast electron heating introduces a non-linear modulation of the metal dielectric constant, provided at first order by $\delta\epsilon({\bf r},t) = \kappa\Delta T_{\rm e}({\bf r},t)$, where $\kappa$ is the complex thermo-derivative coefficient \cite{Marini_2013,marini_faraday}. In turn, accounting also for Kerr nonlinearity arising from the instantaneous conduction hot-electron response, one gets the non-linear polarization ${\bf P}_{\rm NL}({\bf r},t)$ (reported  in Eq. \ref{PolEq}), in which the coefficients $\chi^{(3)}f_{\rm T}$ contain all the time independent terms of the two-temperature model.
  
We calculate  radiation propagation in the medium through the Generalized Nonlinear Schr{\"o}dinger Equation (GNLSE) for the envelope $\psi({\bf r},t)$. Starting from Maxwell's equations coupled with conduction hot-electron dynamics through the non-linear polarization ${\bf P}_{\rm NL}({\bf r},t)$, radiation evolution is described by the inhomogeneous D'Alambert equation $\nabla\times\nabla\times{\bf E} = -\mu_0\partial^2{\bf P}_{\rm NL}/\partial t^2 - (\epsilon(\omega_0)/c^2)\partial^2{\bf E}/\partial t^2$. Thus, adopting the slowly varying envelope approximation (SVEA), one gets to the following evolution equation for the envelope 
\begin{equation}
\partial_z\psi(z,t) = i \frac{k_0}{2}\chi^{(3)} \left[(1-f_{\rm T})|\psi(z,t)|^2 + f_{\rm T} \int_0^{\infty}dt'h_{\rm T}(t')|\psi(z,t-t')|^2\right] \psi(z,t)-\frac{A}{2} \psi(z,t),
\label{GNLSE}
\end{equation}
where $A$ is the Mg attenuation coefficient. Diffraction and dispersion are neglected  owing to the nanometer scale propagation that prevents such effects to play any role. Furthermore, we have assumed that the linear refractive index is $n(\omega_0)\simeq 1$, which is consistent with transmission measurements of Mg at the frequency range of interest. The theoretical results reported above are obtained through a numerical solution of Eq. (\ref{GNLSE}), implemented by a fourth-order Runge-Kutta algorithm complemented with fast Fourier transform.
We assume for $\psi(0, t)$ a Gaussian transform-limited profile with the experimental time duration (59 fs). The Fourier transform $\psi(0, E_{\rm ph})$ is compatible with the average experimental spectral profile (see supplementary information).
The experimental data are compared with the difference of normalized $|\psi(L,E_{\rm ph})|^2$ and $|\psi(0,E_{\rm ph})|^2$ ($\Delta |\psi(L,E_{\rm ph})|^2$).

\section*{Acknowledgements}
We thank L. Giannessi for the useful discussions about the FEL operating system. This project has received funding from the European Union’s Horizon 2020 research and innovation programme Graphene Flagship under grant agreement No 881603 and from PRIN 2017 Project 201795SBA3 – HARVEST.


\newpage
\renewcommand{\thefigure}{S\arabic{figure}}
\setcounter{figure}{0}    
\renewcommand{\theequation}{S\arabic{equation}}

\section{Supporting informations}

\begin{figure}[H]
	\centering
	\includegraphics[width=11cm]{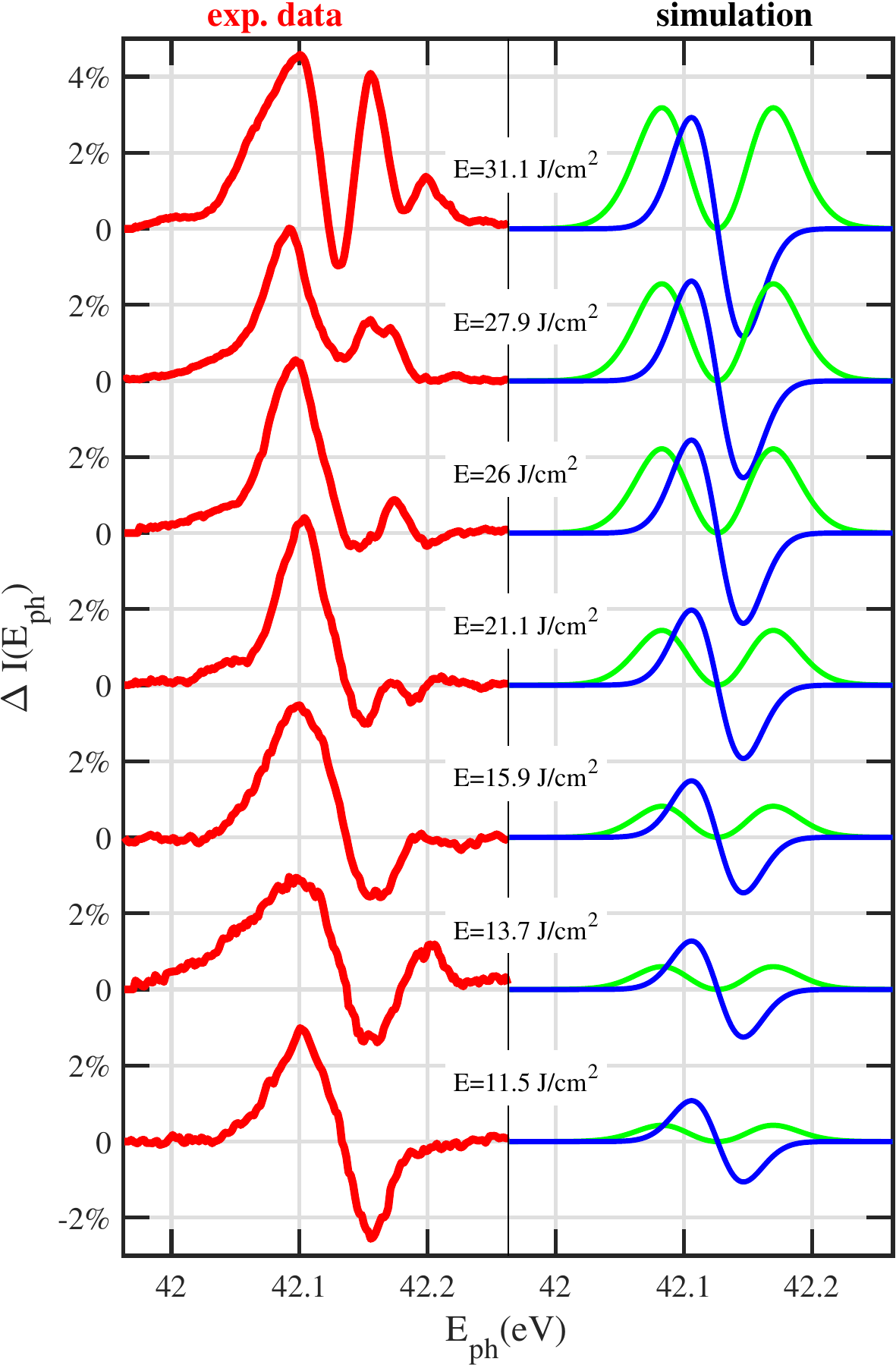} 
	\caption{\textbf{Theoretical simulation of experimental results by SPM and by DTRE.} The experimental data (red lines) reported for different FEL fluences are compared with the relative simulation based on DTRE (blue lines) and SPM (green lines) separately.
		The modification of the electromagnetic wave ($\Delta |\psi(L,\omega)|^2$) is calculated for $L=140$ nm of Mg (see method section). Interestingly, both DTRE and SPM effects are not able to reproduce the experimental results.  
		As shown in Fig. \ref{fig_compare} of the manuscript, 
		the simultaneous action of both the
		effects is necessary to reproduce the experimental data.
	} 	\label{fig_compare_only}
\end{figure} 
\subsection{Single shot FEL spectra}
\begin{figure}[H]
	\centering
	\includegraphics[width=11cm]{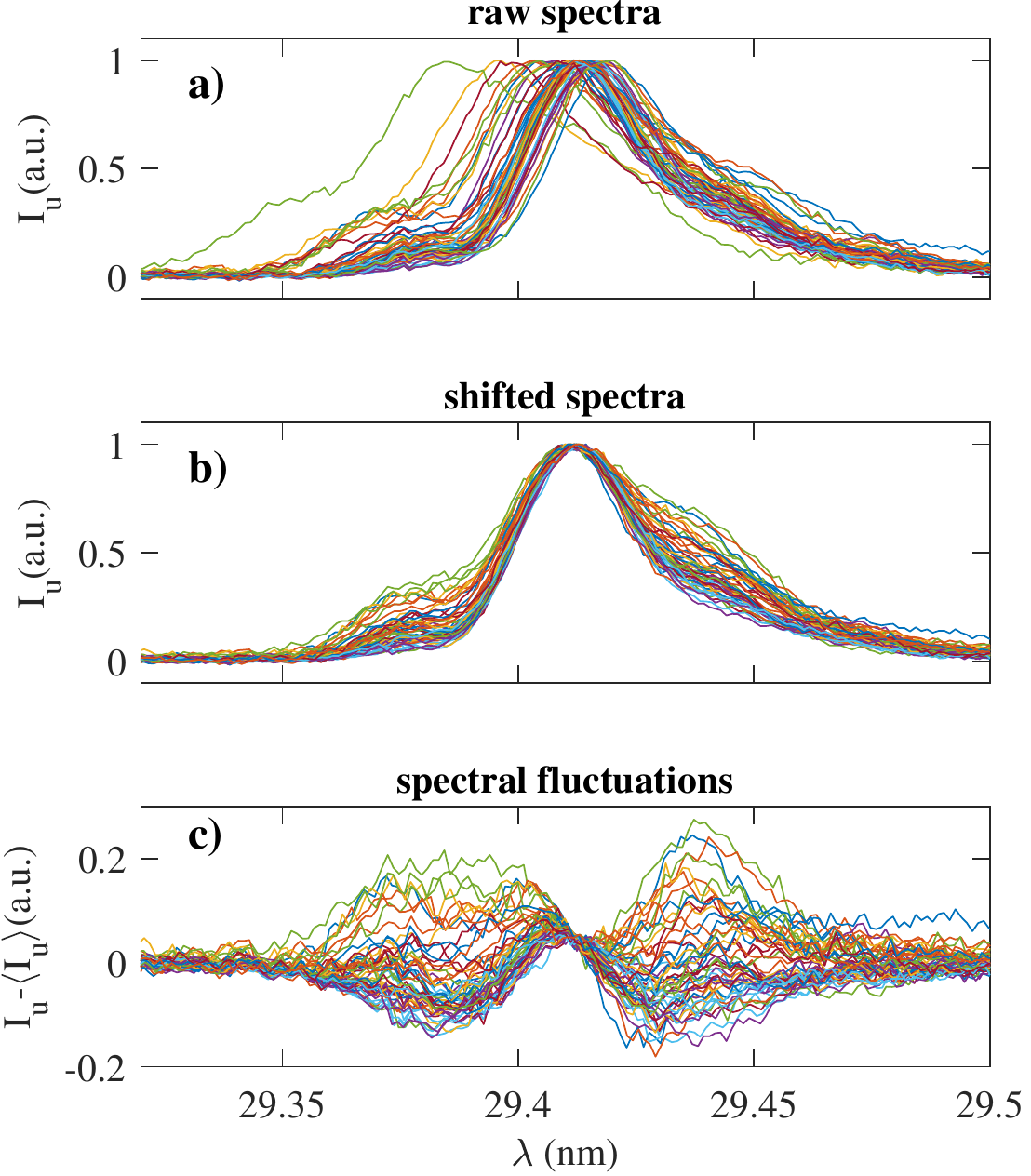} 
	\caption{\textbf{Single FEL shot spectra.} Raw FEL spectra collected by the upstream spectrometer (a) are compared with spectrally shifted spectra (b) and differential spectra (c). 
	} 
	\label{fig_raw_spectra}
\end{figure}
\begin{figure}[H]
	\centering
	\includegraphics[width=11cm]{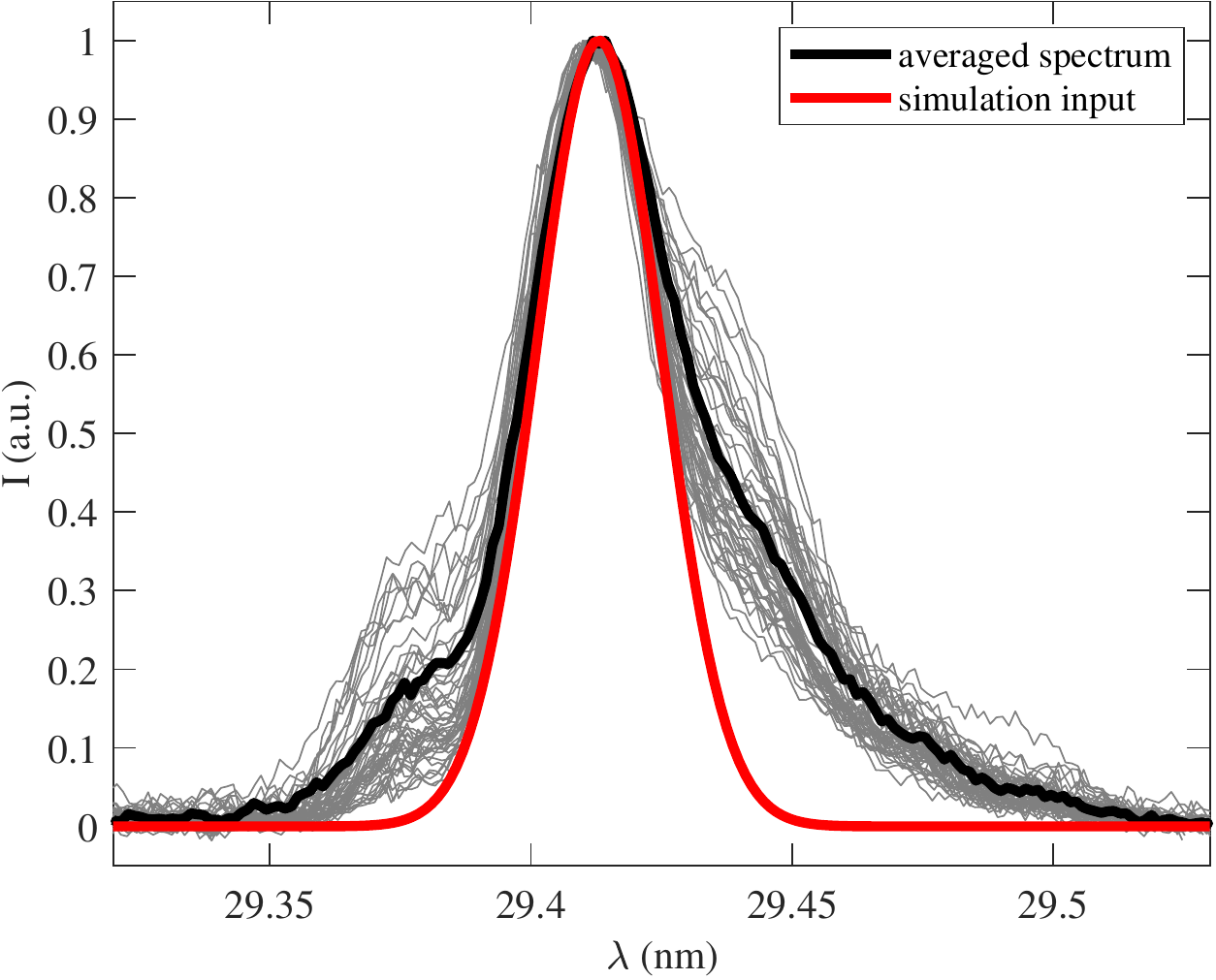} 
	\caption{The grey lines represent the FEL spectra (measured by the upstream spectrometer) for a specific power and the black line is their average. The red line is the spectral profile, obtained as the Fourier transform of a transform limited 59-fs pulse, used in the simulations. 
	} 
	\label{fig_59fs}
\end{figure}

 
Single pulse spectra measured upstream the sample are used as a reference to take into account for FEL fluctuations.
Fig. \ref{fig_raw_spectra} shows indeed that: (i) Single pulse spectra show non negligible random offsets. (ii) Relative shifting single pulse spectra still does not result in a single profile, i.e. lineshape fluctuations as large as 10\% are still present.  
Accordingly, the non-linear self-driven effect is extracted by evaluating the difference of a single downstream spectrum with respect to its upstream reference (shifted to maximize the spectral overlap). 
For what concerns the theoretical modelling of the FEL spectra, we have used a Fourier-transformed 59 fs Gaussian pulse. As shown in Figure \ref{fig_59fs}, two stochastic spectral contributions are typically measured in the single shot spectral profiles. This latter
is indeed typically affected by satellite pulses with random temporal, spatial and spectral shift with respect to the main peak , as discussed in Fig. 12 of Ref \cite{PhysRevX.7.021043}. In our simulation we have not considered the non-linear contribution arising from these weaker and unstable contributions.

\end{document}